\documentclass[twocolumn]{jpsj3}

\usepackage{times}
\usepackage{graphicx}

\title{
Quadrupole Susceptibility of Gd-Based Filled Skutterudite Compounds
}

\author{
Fumiaki {\sc Niikura} and Takashi {\sc Hotta}
}

\inst{
Department of Physics, Tokyo Metropolitan University,
Hachioji, Tokyo 192-0397, Japan
}

\recdate{\today}

\abst{
It is shown that quadrupole susceptibility can be detected
in Gd compounds contrary to our textbook knowledge
that Gd$^{3+}$ ion induces pure spin moment
due to the Hund's rules in an $LS$ coupling scheme.
The ground-state multiplet of Gd$^{3+}$
is always characterized by $J$=7/2,
where $J$ denotes total angular momentum,
but in a $j$-$j$ coupling scheme,
one $f$ electron in $j$=7/2 octet carries
quadrupole moment, while other six electrons
fully occupy $j$=5/2 sextet,
where $j$ denotes one-electron total angular momentum.
For realistic values of Coulomb interaction and
spin-orbit coupling,
the ground-state wavefunction is found to contain
significant amount of the $j$-$j$ coupling component.
From the evaluation of quadrupole susceptibility in a simple mean-field
approximation, we point out a possibility to detect the softening
of elastic constant in Gd-based filled skutterudites.
}

\kword{
Quadrupole, Gadolinium, Filled skutterudites, Wigner-Eckart theorem.
}

\begin{document}
\maketitle

%%%%%%%%%%%%%%%%%%%%%%%%%%%%%%%%%%%%%%%%%%%%%%%%%%%%%%%%%%%%%%%%%%%%%%
%   Sec.1  Introduction
%%%%%%%%%%%%%%%%%%%%%%%%%%%%%%%%%%%%%%%%%%%%%%%%%%%%%%%%%%%%%%%%%%%%%%
\section{Introduction}

Recently, multipole phenomena have attracted much attention
in the research field of strongly correlated $f$-electron systems.
\cite{Hotta1, Review2, Review3}
In general, multipole indicates spin-orbital complex degree of freedom,
which is considered to be active in $f$-electron materials
due to the strong spin-orbit coupling of $f$ electrons.
However, when orbital degeneracy is lifted, for instance,
due to the effect of crystal structure with low symmetry,
only spin degree of freedom often remains.
Thus, $f$-electron compounds with high symmetry are
important for the research of multipole phenomena.
In this sense, filled skutterudite compounds LnT$_4$X$_{12}$
(Ln: lanthanide, T: transition metal, X: pnictogen)
with cubic structure of $T_{\rm h}$ point group
have provided us ideal stages for multipole physics.\cite{skut}

Here we pick up Gd-based compounds.
In general, Gd ion takes the trivalent state in compounds,
leading to the $4f^7$ system.
In an $LS$ coupling scheme, due to the Hund's rules,
electron configuration is determined so as to maximize
total spin angular momentum $S$,
indicating $S$=7/2 for Gd$^{3+}$.
Since seven $f$ orbitals are singly occupied,
total orbital angular momentum $L$ is zero in this case.
Then, the total angular momentum $J$ of the ground state
is given by $J$=$S$=7/2.
The ground-state multiplet is generally affected by
a crystalline electric field (CEF) potential,
but the octet of Gd ion is not changed at all,
since the CEF potentials act only on charge, not on spin,
and the octet with $J$=$S$=7/2 and $L$=0 is spherically symmetric state.
Namely, in the $LS$ coupling scheme, there is no CEF splitting
in the ground-state multiplet
and we cannot expect to observe the response of higher-rank multipole
in Gd-based compounds.

Note here that the $LS$ coupling scheme is exact
in the limit of $U \gg \lambda$,
where $U$ denotes the magnitude of Hund's rule interaction and
$\lambda$ is spin-orbit coupling.
In actual materials, $U$ is larger than $\lambda$, but $\lambda/U$
is not equal to zero.
Thus, the effect of a $j$-$j$ coupling scheme should appear,
more or less, in the ground-state wavefunction.
For Gd$^{3+}$ ion, the ground state in the $j$-$j$ coupling scheme
is also characterized by $J$=7/2, but it is composed of
fully occupied $j$=5/2 sextet and one electron in $j$=7/2 octet,
where $j$ indicates one-electron total angular momentum.
Since one $f$ electron in the $j$=7/2 octet possesses
orbital degree of freedom,
we can observe the CEF level splitting
even for Gd compounds in the $j$-$j$ coupling scheme.
For actual rare-earth compounds, $U$ is about a few eV,
while $\lambda$ is several thousand Kelvin,
indicating that $\lambda/U$ is in the order of 0.1.
Namely, the actual materials are in the intermediate situation
between the $LS$ and $j$-$j$ coupling schemes.
Thus, the ground state of Gd compounds includes
the contribution of the $j$-$j$ coupling scheme
and the quadrupole susceptibility is expected to be significant
in contrast to our knowledge on the basis of the $LS$ coupling scheme.

In this paper, it is shown that quadrupole susceptibility
can be detected in Gd compounds,
although it has been simply believed that
Gd$^{3+}$ ion induces pure spin moment due to the Hund's rules.
It is found that the $j$-$j$ coupling component appears
in the ground-state wavefunction for realistic values of
Coulomb interaction and spin-orbit coupling.
With the use of the CEF parameters determined from PrOs$_4$Sb$_{12}$,
\cite{Kohgi,Kuwahara,Goremychkin}
we evaluate the quadrupole susceptibility for
Gd-based filled skutterudite compounds
and discuss the detection of the softening of elastic constant.

The organization of this paper is as follows.
In Sec.~2, we define the local $f$-electron Hamiltonian
and explain the definition of multipole
in the form of one-body operator.
In Sec.~3, we discuss the CEF states and determine the parameters
from the experimental results of PrOs$_4$Sb$_{12}$.
In Sec.~4, we evaluate quadrupole susceptibility
in a simple mean-field theory
and discuss the detection of the elastic constant
for Gd-based filled skutterudites.
Finally, in Sec.~5, we summarize this paper.
Throughout this paper, we use such units as $\hbar$=$k_{\rm B}$=1.

%%%%%%%%%%%%%%%%%%%%%%%%%%%%%%%%%%%%%%%%%%%%%%%%%%%%%%%%%%%%%%%%%%%%%%
%   Sec.2   Local Hamiltonian and Multipole Operator
%%%%%%%%%%%%%%%%%%%%%%%%%%%%%%%%%%%%%%%%%%%%%%%%%%%%%%%%%%%%%%%%%%%%%%
\section{Local Hamiltonian and Multipole Operator}

\subsection{Local Hamiltonian}

First we define the local $f$-electron Hamiltonian as
\begin{equation}
  \label{Hloc}
  H_{\rm loc}=H_{\rm C}+H_{\rm so}+H_{\rm CEF},
\end{equation}
where $H_{\rm C}$ denotes Coulomb interaction term,
$H_{\rm so}$ is a spin-orbit coupling term,
and $H_{\rm CEF}$ indicates crystalline electric field (CEF) potential term.
Among them, $H_{\rm C}$ is given by
\begin{equation}
H_{\rm C}=\sum_{m_1 \sim m_4}\sum_{\mib{i},\sigma,\sigma'}
  I_{m_1m_2,m_3m_4}
  f_{\mib{i}m_1\sigma}^{\dag}f_{\mib{i}m_2\sigma'}^{\dag}
  f_{\mib{i}m_3\sigma'}f_{\mib{i}m_4\sigma},
\end{equation}
where $\sigma$=$+1$ ($-1$) for up (down) spin,
$f_{\mib{i}m\sigma}$ is the annihilation operator for $f$ electron with
spin $\sigma$ and $z$-component $m$ of angular momentum $\ell$=3
at an atomic site $\mib{i}$,
and the Coulomb integral $I_{m_1m_2,m_3m_4}$ is given by
\begin{equation}
   I_{m_1m_2,m_3m_4} = \sum_{k=0}^{6} F^k c_k(m_1,m_4)c_k(m_2,m_3).
\end{equation}
Here $F^k$ indicates the Slater-Condon parameter and
$c_k$ is the Gaunt coefficient.\cite{Slater}
Note that the sum is limited by the Wigner-Eckart theorem to
$k$=0, 2, 4, and 6.
Although the Slater-Condon parameters should be determined for the material
from the experimental results,
here we simply assume the ratio among the Slater-Condon parameters
as physically reasonable value, given by
\begin{equation}
  F^0=10U,~F^2=5U,~F^4=3U,~F^6=U,
\end{equation}
where $U$ denotes the scale of
Hund's rule interaction among $f$ orbitals,
which should be in the order of a few eV.

The spin-orbit coupling term $H_{\rm so}$ is expressed as
\begin{equation}
   H_{\rm so} = \lambda \sum_{\mib{i},m,\sigma,m',\sigma'}
   \zeta_{m,\sigma;m',\sigma'} f_{\mib{i}m\sigma}^{\dag}f_{\mib{i}m'\sigma'},
\end{equation}
where $\lambda$ is the spin-orbit interaction,
and the matrix element $\zeta$ is given by
\begin{eqnarray}
  \begin{array}{l}
    \zeta_{m,\sigma;m,\sigma}=m\sigma/2,\\
    \zeta_{m+\sigma,-\sigma;m,\sigma}=\sqrt{\ell(\ell+1)-m(m+\sigma)}/2,
  \end{array}
\end{eqnarray}
and zero for other cases.

Finally, the CEF term $H_{\rm CEF}$ is expressed as
\begin{equation}
   H_{\rm CEF} = \sum_{\mib{i},m',\sigma} B_{m,m'}
        f_{\mib{i}m \sigma}^{\dag} f_{\mib{i}m' \sigma},
\end{equation}
where $B_{m,m'}$ denotes the CEF potential
for $f$ electrons from the ligand ions,
which is determined from the table of Hutchings
for angular momentum $\ell$=3.\cite{Hutchings}
For the filled skutterudite compounds
with $T_{\rm h}$ symmetry,\cite{Takegahara}
$B_{m,m'}$ is expressed by using three CEF parameters,
$B_4^0$, $B_6^0$, $B_6^2$, as
\begin{eqnarray}
  \begin{array}{l}
    B_{3,3}=B_{-3,-3}=180B_4^0+180B_6^0, \\
    B_{2,2}=B_{-2,-2}=-420B_4^0-1080B_6^0, \\
    B_{1,1}=B_{-1,-1}=60B_4^0+2700B_6^0, \\
    B_{0,0}=360B_4^0-3600B_6^0, \\
    B_{3,-1}=B_{-3,1}=60\sqrt{15}(B_4^0-21B_6^0),\\
    B_{2,-2}=300B_4^0+7560B_6^0,\\
    B_{3,1}=B_{-3,-1}=24\sqrt{15}B_6^2,\\
    B_{2,0}=B_{-2,0}=-48\sqrt{15}B_6^2,\\
    B_{1,-1}=-B_{3,-3}=360B_6^2.
  \end{array}
\end{eqnarray}
Note the relation of $B_{m,m'}$=$B_{m',m}$.
Following the traditional notation,\cite{LLW}
we define $B_4^0$, $B_6^0$, and $B_6^2$ as
\begin{eqnarray}
  \label{eq:CEF0}
  \begin{array}{l}
    B_4^0=Wx/F(4),\\
    B_6^0=W(1-|x|)/F(6),\\
    B_6^2=Wy/F^t(6),
  \end{array}
\end{eqnarray}
where $x$ and $y$ specify the CEF scheme for $T_{\rm h}$ point group,
while $W$ determines the energy scale for the CEF potential.
As for $F(4)$, $F(6)$, and $F^t(6)$, we choose
$F(4)$=15, $F(6)$=180, and $F^t(6)$=24 for $\ell$=3.\cite{Hutchings}

\subsection{Definition of multipole operator}

Next we define multipole as spin-orbital density
in the form of one-body operator from the viewpoint of
multipole expansion of electron density in electromagnetism.
On the basis of this definition of the multipole operator,
we have developed microscopic theories
for multipole-related phenomena.
For instance, octupole ordering in NpO$_2$ has been clarified
by evaluating multipole interaction with the use of
the standard perturbation method
in terms of electron hopping.\cite{Kubo1,Kubo2,Kubo3}
We have also discussed possible multipole states of
filled skutterudites by analyzing multipole susceptibility of
a multiorbital Anderson model based on the $j$-$j$ coupling scheme.
\cite{Hotta2,Hotta3,Hotta4,Hotta5,Hotta6,Hotta7,Hotta8,Hotta9,Hotta10,Hotta11,Hotta12,Hotta13}
Quite recently, a microscopic theory
for multipole ordering from an itinerant picture
has been developed on the basis of a seven orbital
Hubbard model with spin-orbit coupling.\cite{Hotta14}

The multipole operator ${\hat T}$ is expressed
in the second-quantized form as
\begin{equation}
  {\hat T}^{(k)}_{\mib{i},\gamma} = \sum_{q,m\sigma,m'\sigma'}
  G^{(k)}_{\gamma,q}
  T^{(k,q)}_{m\sigma,m'\sigma'}f^{\dag}_{\mib{i}m\sigma}f_{\mib{i}m'\sigma'},
\end{equation}
where $k$ is a rank of multipole,
$q$ denotes an integer between $-k$ and $k$,
$\gamma$ is a label to express $O_{\rm h}$ irreducible representation,
$G^{(k)}_{\gamma,q}$ is the transformation matrix
between spherical and cubic harmonics,
and $T^{(k,q)}_{m\sigma,m\sigma'}$ can be calculated
by using the Wigner-Eckart theorem as \cite{Inui}
\begin{equation}
 \label{Tkq}
 \begin{split}
  T^{(k,q)}_{m\sigma,m'\sigma'}
  &= \sum_{j,\mu,\mu'}
  \frac{\langle j || T^{(k)} || j \rangle}{\sqrt{2j+1}}
  \langle j \mu | j \mu' k q \rangle \\
  &\times 
  \langle j \mu | \ell m s \frac{\sigma}{2} \rangle
  \langle j \mu' | \ell m' s \frac{\sigma'}{2} \rangle.
 \end{split}
\end{equation}
Here $\ell$=3, $s$=1/2, $j$=$\ell$$\pm$$s$,
$\mu$ denotes the $z$-component of $j$,
$\langle j \mu | j' \mu' j'' \mu'' \rangle$ indicates
the Clebsch-Gordan coefficient,
and $\langle j || T^{(k)} || j \rangle$ is
the reduced matrix element for spherical tensor operator,
given by
\begin{equation}
  \label{red}
  \langle j || T^{(k)} || j \rangle=
  \frac{1}{2^k} \sqrt{\frac{(2j+k+1)!}{(2j-k)!}}.
\end{equation}
We note that $k$$\le$$2j$ and the highest rank is $2j$.
Namely, for $f$ electrons, we can treat the multipoles
up to rank 7 in the present definition.

In this paper, we define the multipole operator
in the one-body form on the analogy of the multipole expansion theory
of electromagnetic potential,
but some readers may cast questions about the validity of
the present definition.
Since the multipole operators have been traditionally defined by
the Stevens' operator equivalent in the space of the Hund's-rule multiplet
characterized by the total angular momentum $J$,\cite{StevensOP}
the present definition seem to be different from
such a traditional one at a first glance.
Concerning this point, we provide detailed comments in the following.
Hereafter the Stevens' operator equivalent is simply called
the Stevens operator.

First of all, we reply to the above question in a simple manner.
In one word, the present definition of multipole operator
is related to the Stevens operator through the Wigner-Eckart theorem as
\begin{equation}
   \label{w-e}
  \begin{split}
  & \langle n, J, M | {\hat T}^{(k)}_{\gamma} |n, J, M' \rangle \\
  & =c^{(k)}_{n, J}(U, \lambda) \sum_q G^{(k)}_{\gamma,q} O_k^q(J; M, M'),
\end{split}
\end{equation}
where
$n$ is the local $f$-electron number,
$J$ is the total angular momentum of the ground state multiplet,
$M$ is the $z$-component of $J$,
$|n, J, M \rangle$ denotes the degenerate ground states
of $H_{\rm C}+H_{\rm so}$,
$c$ denotes a proportional coefficient depending on 
$n$, $J$, $k$, $U$, and $\lambda$,
and $O_k^q(J; M, M')$ is the matrix element of
Stevens operator ${\hat O}_k^q$,
given by
\begin{equation}
 \label{Stevens}
  O_k^q(J; M, M')=
  \frac{\langle J || T^{(k)} || J \rangle}{\sqrt{2J+1}}
  \langle J M | J M' k q \rangle.
\end{equation}
Here the reduced matrix element is given by eq.~(\ref{red}),
in which $j$ is replaced with $J$.
We can numerically calculate all the matrix elements of
the Stevens operator for arbitral numbers of $J$ and $k$,
but for some low-order multipoles,
it is convenient to use the symmetrized polynomials of $J_{\alpha}$,
\cite{Shiina1,Shiina2}
where $J_{\alpha}$ ($\alpha$=$x$, $y$, and $z$) is the $\alpha$
component of total angular momentum $J$.

%%%%%%%%%%%%%%%%%%%%%%% Fig. 1 %%%%%%%%%%%%%%%%%%%%%%%%%%%
\begin{figure}[t]
\centering
\includegraphics[width=8.0truecm]{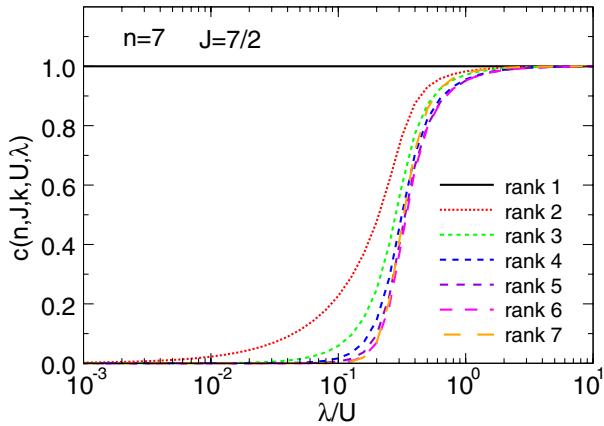}
\caption{(Color online)
Coefficients $c$ of the multipole operators for $n$=7
obtained from the diagonalization of $H_{\rm C}+H_{\rm so}$.
}
\end{figure}
%%%%%%%%%%%%%%%%%%%%%%%%%%%%%%%%%%%%%%%%%%%%%%%%%%%%%%%%%%

In Fig.~1, we depict the coefficients $c$ for the case of $n$=7,
corresponding to Gd$^{3+}$ ion.
We note that the coefficients do not depend only on
the ratio of $\lambda/U$,
but even if the values of $U$ and $\lambda$ have been
individually changed so as to keep the ratio,
we have not found significant difference in the results.
Note also that in the present case of $n$=7,
the ground state multiplet is characterized by $J$=7/2 and
the highest multipole is rank 7 in any case.
As naively expected, we find that only dipole moment remains
in the limit of the $LS$ coupling scheme for large $U$,
since the total angular momentum $J$ of the ground-state multipole
is purely composed of seven spins.
In the limit of large $\lambda$,
all the coefficients converge to unity,
since the ground state is essentially expressed by
one electron in the $j$=7/2 octet
and the one-body multipole expression perfectly agrees with
the Stevens operator in the one-electron case.
It is interesting to mention that this fact
is also related to Yb$^{3+}$ with
one hole in the $j$=7/2 octet.

It should be noted that around at $\lambda/U$=0.1$\sim$0.2,
corresponding to the region for realistic values
of actual materials,
the coefficients for multipoles other than dipole rapidly
increase,
although the magnitude of the coefficient depends on
individual values of $U$ and $\lambda$.
In particular, it is impressive that the coefficient for
quadrupole seems to be larger than we have naively expected,
since the realistic situation with $\lambda/U$=0.1$\sim$0.2
has been believed to be well described by the $LS$ coupling scheme.
This result strongly suggests that quadrupole degree of freedom
cannot be simply neglected in Gd compounds.
The extent of the effect of quadrupole will be discussed
in the next section.

From a mathematical viewpoint, the relation eq.~(\ref{w-e}) just
expresses the Wigner-Eckart theorem, but it contains physically
important message concerning the multipole.
Since the $J$-multiplet of $f^n$-electron system is given
by the appropriate superposition of $n$-body states,
the Stevens operator is considered to be given by
the expression of multipole in the many-body form,
in sharp contrast to the present definition of one-body form.
However, for $f^n$-electron systems, eq.~(\ref{w-e}) means that
the expectation value of the multipole operator in the present definition
is in proportion to the Stevens operator and the effect of interactions
is included only in the proportional coefficient $c$.
When we impose the orthonormal condition on the multipole operator,
the difference in the normalization is absorbed
in the proportional coefficient.
Namely, the multipole in the present definition is equivalent to
the Stevens operator, except for the proportional coefficient.
Thus, it is possible to use one of the definitions of multipole,
one-body or many-body form, depending on the problem,
but we point out some advantages to use the multipole
in the one-body form.

The Stevens operator is considered to be suitable for localized
situation, which is well described by the $LS$ coupling scheme.
First we construct the Hund's-rule ground state characterized
by spin moment $S$ and angular momentum $L$ for $f^n$-electron systems.
Then, we include the effect of spin-orbit coupling
to form the multiplet characterized by total angular momentum $J$,
which is given by $J$=$|L$$-$$S|$ and $J$=$L$+$S$ for $n$$<$7 and
$n$$>$7, respectively.
In such a case, it seems to be natural to define the multipole
by the Stevens operator given by eq.~(\ref{Stevens}).
However, it is difficult to consider itinerant situation
in the $LS$ coupling scheme, since $J$ is changed by charge fluctuations.
In contrast, the present definition of multipole can be used both for
itinerant and localized cases, since it is defined as
the one-body density operator.
This is one of advantages of the present definition of multipole.

When we define multipole in the one-body density operator,
the effect of Coulomb interaction and spin-orbit coupling
is automatically included in the proportional coefficient.
Thus, in the present definition of multipoles,
we can discuss the relative change of multipole moment
in the wide range of Coulomb interaction and spin-orbit coupling
from $\lambda/U$=0 ($LS$ coupling scheme) to
$\lambda/U$=$\infty$ ($j$-$j$ coupling scheme).
This is another advantage of the one-body definition of multipole.
It is true that from a symmetry viewpoint, we can continue to use
the Stevens operator even if we deviate from the $LS$ coupling scheme,
since the ground state multiplet is always characterized
by the same total angular momentum when we change $U$ and/or $\lambda$.
However, the coefficient $c$ in eq.~(\ref{w-e}) is changed by
the interactions, but this point is $not$ included
in the Stevens operator.
This is easily understood when we recall the CEF potential
for the case of $n$=2, in which the ground state multiplet is
characterized by $J$=4.
The CEF potentials are expressed by using the fourth- and
sixth-order Stevens operators, but the sixth-order one vanishes
in the $j$-$j$ coupling regime,
since the CEF term is originally given by the sum of one-electron
electrostatic potential and the sixth-order terms
do not appear for one electron in the $j$=5/2 states
which are accommodated in the limit of large $\lambda$.

Here we should mention that there is a disadvantage concerning
the limitation of the rank of multipole in the present definition
of the multipole in the one-body form.
As mentioned in the previous subsection, we can treat the multipoles
up to rank 7 in the present multipole definition.
In order to treat higher multipoles with $k$$\ge$8, it is necessary
to define the multipole in the many-body operator form,
since the Stevens operator with the many-body form can treat
all possible multipoles up to rank $2J$
in the multiplet characterized by $J$.
Nevertheless we believe that the present multipole definition
with one-body form keeps an advantage, since in actuality,
it is very rare to treat higher-order multipoles with $k$$\ge$8.

%%%%%%%%%%%%%%%%%%%%%%%%%%%%%%%%%%%%%%%%%%%%%%%%%%%%%%%%%%%%%%%%%%%%%%
%   Sec.3   CEF states
%%%%%%%%%%%%%%%%%%%%%%%%%%%%%%%%%%%%%%%%%%%%%%%%%%%%%%%%%%%%%%%%%%%%%%
\section{Crystalline Electric Field States}

\subsection{Effective Hamiltonian}

The local CEF state can be determined by the diagonalization of
the local Hamiltonian eq.~(\ref{Hloc}),
but in order to understand how the CEF states are changed
due to the competition between the Coulomb interaction
and spin-orbit coupling, i.e.,
the $LS$ and $j$-$j$ coupling schemes,
it is useful to derive the effective Hamiltonian
$H_{\rm eff}$ from eq.~(\ref{Hloc})
with the use of the Stevens operator eq.~(\ref{Stevens}).
We express $H_{\rm eff}$ as
\begin{equation}
\label{eq:eff}
  \begin{split}
  H_{\rm eff}(n,J) &= B_4^0(n,J)({\hat O}_4^0+5{\hat O}_4^4) \\
   &+ B_6^0(n,J)({\hat O}_6^0-21{\hat O}_6^4) \\
   &+ B_6^2(n,J)({\hat O}_6^2-{\hat O}_6^6),
  \end{split}
\end{equation}
where ${\hat O}_k^q$ is the Stevens operator
of which matrix elements for $n$ and $J$
are given by eq.~(\ref{Stevens})
and $B_k^q(n,J)$ denotes the CEF parameter
for the multiplet specified by $n$ and $J$.

The effects of $U$ and $\lambda$ appear in the CEF parameters,
$B_4^0(n,J)$, $B_6^0(n,J)$, and $B_6^2(n,J)$,
which are expressed by
\begin{eqnarray}
  \label{eq:CEF1}
  \begin{array}{l}
   B_4^0(n,J) =k_{4}(n,J) B_4^0,\\
   B_6^0(n,J) =k_{6}(n,J) B_6^0,\\
   B_6^2(n,J) =k_{6}(n,J) B_6^2,
  \end{array}
\end{eqnarray}
where $B_4^0$, $B_6^0$, and $B_6^2$ for one $f$ electron with $\ell$=3
are given by eq.~(\ref{eq:CEF0})
and $k_4(n,J)$ and $k_6(n,J)$ are given by
\begin{equation}
   \label{eq:CEF3}
   k_4(n,J)=\beta^{(n)}_{J}/\beta_{\ell},~
   k_6(n,J)=\gamma^{(n)}_{J}/\gamma_{\ell},
\end{equation}
respectively.
Here $\beta^{(n)}_{J}$ and $\gamma^{(n)}_{J}$ are the so-called
Stevens factors, which are coefficients
appearing in the method of Stevens' operator equivalent.\cite{StevensOP}
Note that in general, $\beta^{(n)}_{J}$ and $\gamma^{(n)}_{J}$
depend on the Coulomb interaction and spin-orbit coupling,
since they are determined by the nature of the ground-state multiplet
specified by $J$.

For one $f$ electron with $\ell$=3,
$\beta_{\ell}$ and $\gamma_{\ell}$ are given by
\begin{equation}
  \beta_{\ell}=\frac{2}{11\cdot45},~
  \gamma_{\ell}=-\frac{4}{9\cdot13\cdot33},
\end{equation}
respectively.
For the ground state of $n$=7 and $J$=7/2,
in the limit of $U$=$\infty$,
i.e., in the $LS$ coupling scheme,
we obtain $k^{LS}_4(n,J)$=$k^{LS}_6(n,J)$=0,
suggesting that the CEF potentials do not work
for $f^7$ states with $J$=$S$=7/2 in the $LS$ coupling scheme.
On the other hand, in the limit of $\lambda$=$\infty$,
i.e., in the $j$-$j$ coupling scheme,
we obtain $k^{j-j}_4(n,J)$=3/7 and $k^{j-j}_6(n,J)$=1/7
from the Stevens factors ~\cite{StevensOP}
\begin{equation}
\beta^{(7)}_{7/2}=\frac{2}{15\cdot77},~
\gamma^{(7)}_{7/2}=-\frac{4}{13\cdot33\cdot63}.
\end{equation}
Note that the absolute values of $\beta^{(7)}_{7/2}$ and $\gamma^{(7)}_{7/2}$
in the $j$-$j$ coupling scheme are equal to
those for Yb$^{3+}$ ion with one hole in $j$=7/2 octet,
although the signs are just inverted.

%%%%%%%%%%%%%%%%%%%%%%% Fig. 2 %%%%%%%%%%%%%%%%%%%%%%%%%%%
\begin{figure}[t]
\centering
\includegraphics[width=8.0truecm]{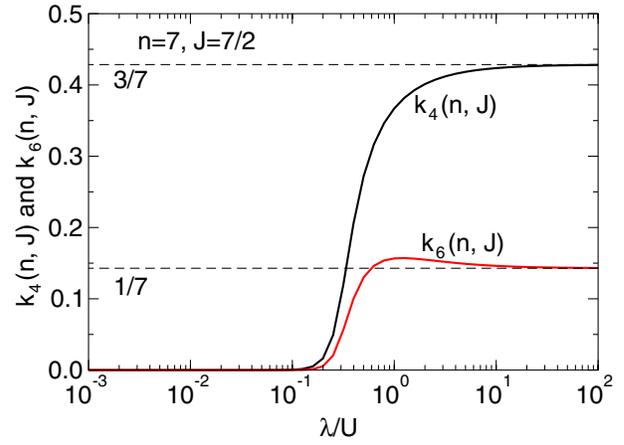}
\caption{(Color online)
Coefficients $k_4(n,J)$ and $k_6(n,J)$ for $n$=7 and $J$=7/2
obtained from the diagonalization of $H_{\rm C}+H_{\rm so}$.
}
\end{figure}
%%%%%%%%%%%%%%%%%%%%%%%%%%%%%%%%%%%%%%%%%%%%%%%%%%%%%%%%%%

For the intermediate coupling region in which both $U$ and $\lambda$
are finite, we numerically evaluate $k_4(n,J)$ and $k_6(n,J)$
by deriving $H_{\rm eff}$ from the local Hamiltonian $H_{\rm loc}$.
By using the ground state $|n,J,M\rangle$ of $H_{\rm so}+H_{\rm C}$,
we evaluate the matrix elements of the effective Hamiltonian as
\begin{equation}
  \label{HS2}
  H_{\rm eff}(n,J;M,M')
  =\langle n,J,M | H_{\rm CEF} | n,J,M' \rangle,
\end{equation}
where we assume that $|W|$ is
much smaller than both $U$ and $\lambda$.
Since the matrix elements of the Stevens operator
have been already listed for each value of $J$,
we can numerically obtain $B_4^0(n,J)$, $B_6^0(n,J)$, and $B_6^2(n,J)$
from eq.~(\ref{HS2}) for any values of $U$ and $\lambda$
for a given value of $n$.

In Fig.~2, we show $k_4(n,J)$ and $k_6(n,J)$ for $n$=7 and $J$=7/2
as functions of $\lambda/U$.
In order to draw this figure,
we numerically evaluate the matrix element eq.~(\ref{HS2})
for $|W|$=$10^{-4}$ eV.
Concerning $U$ and $\lambda$,
for $\lambda/U<1$, we fix $\lambda$ as $\lambda$=1.0 and increase $U$,
while for $\lambda/U>1$, we fix $U$ as $U$=1.0 and increase $\lambda$.
As mentioned in Fig.~1,
we have not found significant difference in the results,
even when we individually change the values of $U$ and $\lambda$
by keeping the ratio.

As mentioned above, we obtain $k_4(n,J)$=$k_6(n,J)$=0
for small $\lambda/U$, while we find $k_4(n,J)$=$3k_6(n,J)$=3/7
for large $\lambda/U$.
In the region of $\lambda/U$=0.1$\sim$1.0, we observe
rapid increases of both $k_4(n,J)$ and $k_6(n,J)$ from zeros 
to the values of the $j$-$j$ coupling limit.
In this sense, for actual values of $\lambda$ and $U$,
we expect the appearance of the CEF effect even for the case of $n$=7.
The quantitative argument of the CEF potentials will be discussed in
the next subsection.

Finally, we note the difference in the increasing behavior
of $k_4(n,J)$ and $k_6(n,J)$.
Although $k_4(n,J)$ monotonically increases from zero to 3/7,
$k_6(n,J)$ does not change monotonically, but
it forms a peak around at $\lambda/U$$\approx$1.0,
In particular, $k_6(n,J)$ is slightly larger than 1/7,
which is the value of the $j$-$j$ coupling limit.
In the region of $\lambda/U$=0.1$\sim$1.0,
the effect of $k_6(n,J)$ is
relatively larger than that of $k_4(n,J)$.

\subsection{CEF parameters}

Before proceeding to the evaluation of quadrupole susceptibility,
it is necessary to determine the CEF parameters in $H_{\rm CEF}$.
For the purpose, as typical material,
we consider filled skutterudite compounds.\cite{skut}
As for the CEF parameters of PrOs$_4$Sb$_{12}$,
it has been confirmed that
the ground state is $\Gamma_1^+$ singlet state
and the first CEF excited state is $\Gamma_{4}^{+(2)}$
triplet with small excitation energy of about 10 K.
\cite{Kohgi,Kuwahara,Goremychkin}
Note that since filled skutterudite compounds crystallize
in the cubic structure with $T_{\rm h}$ symmetry,
two triplet states of $\Gamma_{4}^{+}$ and $\Gamma_5^{+}$
in $O_{\rm h}$ symmetry are mixed with each other,
leading to two $\Gamma_{4}^{+}$ states.\cite{Takegahara}

%%%%%%%%%%%%%%%%%%%%%%% Fig. 3 %%%%%%%%%%%%%%%%%%%%%%%%%%%
\begin{figure}[t]
\centering
\includegraphics[width=8.0truecm]{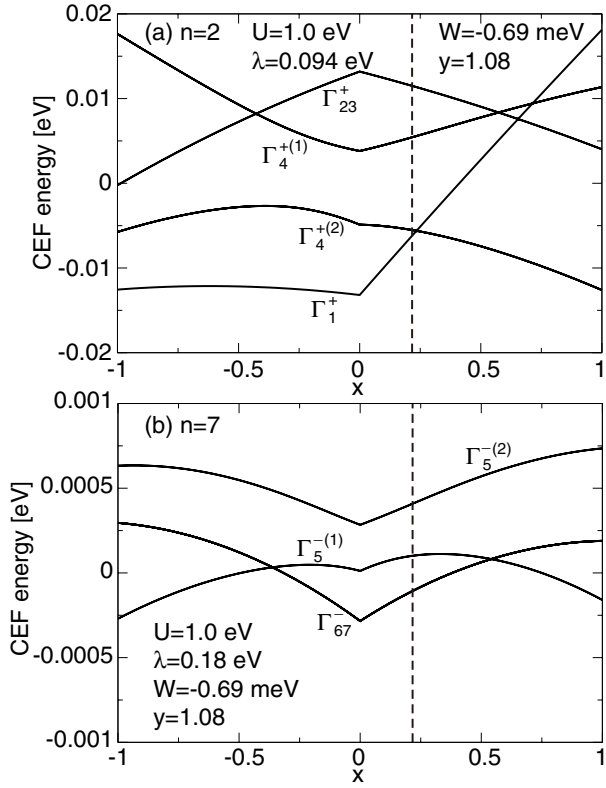}
\caption{CEF energy schemes for (a) $n$=2 and (b) $n$=7.
As for the details of the parameters, see the main text.
Here we note that we set $W$=$-0.69$ meV and $y$=1.08.
}
\end{figure}
%%%%%%%%%%%%%%%%%%%%%%%%%%%%%%%%%%%%%%%%%%%%%%%%%%%%%%%%%%

In order to reproduce the CEF energy scheme which has been
experimentally determined,\cite{Kohgi,Kuwahara,Goremychkin}
we perform the optimization of $W$, $x$, and $y$
in $H_{\rm CEF}$ of eq.~(\ref{Hloc}) for $U$=1.0 eV.
Note that $\lambda$ is set as the value which has been
experimentally found.\cite{spin-orbit}
For Pr$^{3+}$ ion, we use $\lambda$=0.094 eV.
After the optimization, we find
$W$=$-$0.69 meV, $x$=0.22, and $y$=1.08.
%
%%%%%%%%%%%%%%%%%%%%%%%%%%%%%%%%%%%%%%%%%%%%%%%%%%%%%%%%%%%%%%%%%%%%%
%
We note that these CEF parameters determine
the CEF potentials for one $f$-electron state
of eq.~(\ref{eq:CEF0}),
{\it not} for $f^2$-states of $J$=4 of
eqs.~(\ref{eq:eff}) and (\ref{eq:CEF1}).
In Fig.~3(a), we show the CEF energies vs. $x$
for $W$=$-$0.69 meV and $y$=1.08.
The vertical broken line denotes the position of $x$=0.22,
at which we find $E(\Gamma_{4}^{+(2)})-E(\Gamma_{1}^{+})$=7.9K,
$E(\Gamma_{4}^{+(1)})-E(\Gamma_{1}^{+})$=134.9K,
and $E(\Gamma_{23}^{+})-E(\Gamma_{1}^{+})$=205.4K,
where $E(\Gamma_\gamma)$ denotes the eigenenergy
of the CEF state characterized by the irreducible representation
of $\Gamma_{\gamma}$.
These excitation energies reproduce well
the experimental CEF energy scheme
of PrOs$_4$Sb$_{12}$.\cite{Kohgi,Kuwahara,Goremychkin}

Here it is instructive to consider the CEF potentials
in eq.~(\ref{eq:CEF1}) of the effective model for $n$=2 and $J$=4.
After some algebraic calculations with the use of
the above optimized values of $W$, $x$, and $y$,
we obtain $B_4^0(n=2,J=4)$=$2.29 \times 10^{-2}$K,
$B_6^0(n=2,J=4)$=$1.33 \times 10^{-3}$K,
and $B_6^2(n=2,J=4)$=$1.38 \times 10^{-2}$K.
In the analysis of the experimental results,\cite{Kohgi}
the CEF potentials for the states of $n$=2 and $J$=4
 have been found to be
$A_4$=$2.37 \times 10^{-2}$K,
$A_6$=$1.32 \times 10^{-3}$K,
and $A_6^t$=$1.08 \times 10^{-2}$K,
where $A_4$, $A_6$, and $A_6^t$ used in Ref.~\citen{Kohgi}
denote $B_4^0(n=2,J=4)$, $B_6^0(n=2,J=4)$, and
$B_6^2(n=2,J=4)$ in our notation, respectively.
The CEF potentials for the states of $n$=2 and $J$=4 obtained
by considering both finite spin-orbit coupling and Coulomb interactions
are totally similar to those obtained in the $LS$ coupling scheme
in which the Coulomb interactions are assumed to be infinite,
although there exist deviations between two groups of the CEF potentials
due to the effect of finite Coulomb interactions.
This fact indicates that
the CEF states of $n$=2 are well described by the $LS$ coupling scheme.
%
%%%%%%%%%%%%%%%%%%%%%%%%%%%%%%%%%%%%%%%%%%%%%%%%%%%%%%%%%%%%%%%%

Next we show the CEF energy scheme for Gd-based filled
skutterudites by using the same CEF parameters.
Since the CEF potentials work on $f$ orbital,
as easily understood from $H_{\rm CEF}$,
there is no reason to change the CEF parameters
when we replace the rare-earth ion with another one
in the same environment due to the same ligand ions.
Note that $\lambda$ is determined from the experimental value
and we set $\lambda$=0.18 eV for Gd ion.
In Fig.~3(b), we show the CEF energies vs. $T$ for $n$=7
with the use of the same parameters as those in Fig.~3(a),
except for the spin-orbit coupling.
Note that the magnitude of the CEF splitting is
obviously smaller than that of Fig.~3(a),
but we observe the CEF excitation in the order of a few Kelvin.
In the present parameters,
we find that at $x$=0.22, the ground state is
$\Gamma_{67}^-$ quartet and the excited states are
two types of $\Gamma_5^-$ doublets.
Note that two different doublets $\Gamma_6^-$ and $\Gamma_7^-$
in $O_{\rm h}$ symmetry are mixed with each other,
leading to two $\Gamma_{5}^{-}$ doublet states.
Note also that $\Gamma_8^-$ quartet in $O_{\rm h}$ symmetry
is expressed by $\Gamma_{67}^-$ in $T_{\rm h}$ symmetry.
\cite{Takegahara}

In the $LS$ coupling scheme, there is no CEF splitting
for the case of $n$=7, since the octet of $J$=7/2
is composed of $S$=7/2 and $L$=0.
However, as emphasized in the discussion of multipole operator,
the actual situation is in the middle of
the $LS$ and $j$-$j$ coupling schemes.
Thus, in contrast to our knowledge on the basis of the $LS$
coupling scheme, we observe the CEF splitting
due to the effect of the $j$-$j$ coupling component in the wavefunction.
This point will be discussed again for the quadrupole susceptibility.

%%%%%%%%%%%%%%%%%%%%%%%%%%%%%%%%%%%%%%%%%%%%%%%%%%%%%%%%%%%%%%%%%%%%%%
%   Sec.4   quadrupole susceptibility
%%%%%%%%%%%%%%%%%%%%%%%%%%%%%%%%%%%%%%%%%%%%%%%%%%%%%%%%%%%%%%%%%%%%%%
\section{Quadrupole Susceptibility}

Now we discuss the quadrupole susceptibility,
which can be detected by the measurement of elastic constant $C$.
In general, we consider the strain fields $\epsilon_u$ and $\epsilon_v$
which are coupled with quadrupoles
${\hat Q}_{u\mib{i}}$=${\hat T}^{(2)}_{\mib{i},3z^2-r^2}$
and
${\hat Q}_{v\mib{i}}$=${\hat T}^{(2)}_{\mib{i},x^2-y^2}$,
respectively.
In the notation of the Stevens operator,
these are expressed by
$(2J_z^2-J_x^2-J_y^2)/2$
and
$\sqrt{3}(J_x^2-J_y^2)/2$,
respectively.\cite{Shiina1,Shiina2}

By following the standard formalism
to calculate the elastic constant, we add the coupling term
between strain and quadrupole to $H_{\rm loc}$, expressed as
\begin{equation}
  H=H_{\rm loc}+H_{\rm QS}+H_{\rm QQ},
\end{equation}
where $H_{\rm QS}$ and $H_{\rm QQ}$ denote quadrupole-strain interaction
and inter-site quadrupole interaction terms, respectively.
We express them as
\begin{equation}
  H_{\rm QS}=- g \sum_{\mib{i}} ({\hat Q}_{u\mib{i}}
\epsilon_u+{\hat Q}_{v\mib{i}} \epsilon_v),
\end{equation}
and
\begin{equation}
  H_{\rm QQ}=-g' \sum_{\langle \mib{i},\mib{j} \rangle}
  ({\hat Q}_{u\mib{i}} {\hat Q}_{u\mib{j}}
  +{\hat Q}_{v\mib{i}} {\hat Q}_{v\mib{j}}),
\end{equation}
where $g$ and $g'$, respectively, denote quadrupole-strain and
inter-site quadrupole couplings
and $\langle \mib{i},\mib{j} \rangle$ indicates the pair of
nearest neighbor atomic sites.

The elastic constant $C$ is usually expressed with the use of
quadrupole susceptibility $\chi_Q$ as
$C$=$-N g^2 \chi_Q$,
where $N$ denotes the number of rare-earth ion in the unit volume.
By considering only the strain field of $\epsilon_v$,
we estimate $\chi_Q$ in a simple mean-field approximation,
in which we express $H_{\rm QQ}^{\rm MF}$ as
\begin{equation}
  H_{\rm QQ}^{\rm MF}=
  -g' \sum_{\langle \mib{i},\mib{j} \rangle} {\hat Q}_{v\mib{i}}
\langle {\hat Q}_{v\mib{j}} \rangle,
\end{equation}
where $\langle \cdots \rangle$ denotes the operation to take thermal average.
Then, we obtain $\chi_Q$ in the mean-field approximation as
\begin{equation}
 \chi_Q= \frac{\chi} {1-g' \chi},
\end{equation}
where $\chi$ is the on-site quadrupole susceptibility
determined from $H_{\rm loc}$, given by
\begin{equation}
   \chi = \frac{1}{Z}
   \sum_{n,m} \frac{e^{-E_n/T}-e^{-E_m/T}}{E_m-E_n}
   |\langle n | [{\hat Q}_v - \langle {\hat Q}_v \rangle] | m \rangle |^2.
\end{equation}
Here we suppress the site dependence of ${\hat Q}_{v\mib{i}}$,
$Z$ is the partition function given by $Z$=$\sum_n e^{-E_n/T}$,
$E_n$ denotes the eigenenergy of the $n$-th eigenstate
$|n \rangle$ of $H_{\rm loc}$,
and $\langle {\hat Q} \rangle$=$\sum_n e^{-E_n/T}
\langle n |{\hat Q} | n \rangle/Z$.

%%%%%%%%%%%%%%%%%%%%%%% Fig. 4 %%%%%%%%%%%%%%%%%%%%%%%%%%%
\begin{figure}[t]
\centering
\includegraphics[width=8.0truecm]{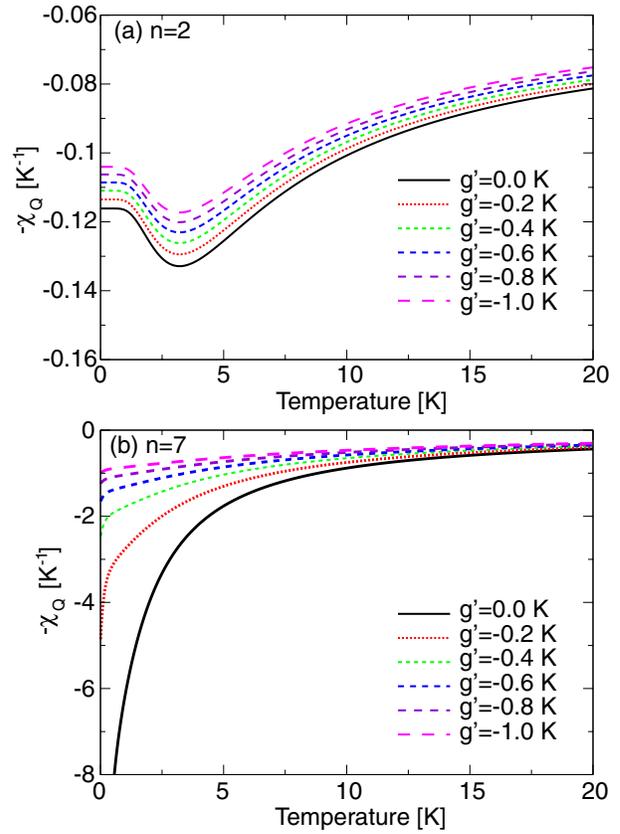}
\caption{(Color online)
Quadrupole susceptibility vs. temperature $T$ for
(a) $n$=2 and (b) $n$=7.
As for the details of the parameters, see the main text.
Note here that the unit of $g'$ is Kelvin.
}
\end{figure}
%%%%%%%%%%%%%%%%%%%%%%%%%%%%%%%%%%%%%%%%%%%%%%%%%%%%%%%%%%

In Fig.~4(a), we depict $-\chi_Q$ vs. $T$ for $n$=2 corresponding to
PrOs$_4$Sb$_{12}$ for several values of $g'$
with the use of $U$=1.0 eV, $\lambda$=0.094 eV (experimental value),
and the CEF parameters determined in the previous subsection.
For $g'$=0,
we find a minimum in $-\chi_Q$ at $T$$\approx$$3$K
due to the suppression of the Curie term.
Then, at low enough temperatures,
the quadrupole susceptibility is dominated by
the Van Vleck term due to the off-diagonal term stemming
from the quasi-degenerate $\Gamma_1$ singlet and
$\Gamma_4^{(2)}$ triplet,\cite{Goto1,Goto2}
which are characteristic to PrOs$_4$Sb$_{12}$.
When we increase the value of $g'$ up to 1 K as shown in Fig.~4(a),
we do not find significant changes in the temperature dependence,
although the absolute value is slightly suppressed.

Here we remark that the characteristic behavior of $-\chi_Q$ for $g'$=0
has been obtained in the analysis of the experimental results
in the $LS$ coupling scheme.\cite{Goto1,Goto2}
The experimental results on the elastic constant for PrOs$_4$Sb$_{12}$
have been well explained by such calculation results.
Here it is emphasized that we correctly reproduce the previous results
by using the definition of
multipole in the one-electron density operator form.
Note, however, that the absolute value of $-\chi_Q$ for $g'$=0
is different from that in the previous result,
although the shape of the temperature dependence agrees quite well
with the previous one.
An important origin of such difference
is due to the difference in the wavefunction.
Another origin is due to the normalization of the multipole operator,
which will be commented later.
In any case, for the purpose to reproduce the experimental results
with the calculation one, we adjust the value of $g$,
which is just a fitting parameter, since it is not determined
solely from the theoretical calculations in the present framework.
Thus, in this paper, we do not take care of the difference
in the absolute values between the present $\chi_Q$ and
the previous results.
However, it is meaningful to consider the relative difference
between the results for $n$=2 and $n$=7
in the same calculation framework.

In Fig.~4(b), we depict $-\chi_Q$ vs. $T$ for $n$=7 corresponding to
Gd-based filled skutterudite compounds for several values of $g'$
with the use of $U$=1.0 eV, $\lambda$=0.18 eV (experimental value),
and the same CEF parameters as those for PrOr$_4$Sb$_{12}$.
The results are similar to those for Ce compounds with $\Gamma_8$
quartet ground state.
This point is intuitively understood from the similarity between
Ce$^{3+}$ ion with one electron in the $j$=5/2 sextet and
Gd$^{3+}$ ion with one electron in the $j$=7/2 octet in the limit
of the $j$-$j$ coupling scheme,
when quartet ground state appears due to the CEF potentials.

Note that as easily understood from the difference in the vertical
scales between Figs.~4(a) and 4(b),
the absolute value of the quadrupole susceptibility
for $n$=7 is larger than that for $n$=2,
mainly due to the contribution of the Curie term.
When we increase $g'$ in the order of 1 K,
the magnitude of the quadrupole susceptibility is totally suppressed,
but it is still possible to observe the softening in the elastic constant,
which is determined by the quadrupole susceptibility.
For the case of PrOs$_4$Sb$_{12}$, the fitting to the
experimental results has been done for $g'$=0,\cite{Goto1,Goto2}
but at the present stage,
we cannot determine the magnitude of $g'$
in order to discuss the absolute value of $-\chi_Q$
of Gd-based filled skutterudites.
An important message which we emphasize here is
that it is possible to observe the softening of elastic constant
even in Gd-based filled skutterudite compounds
contrary to our knowledge on the basis of
the $LS$ coupling scheme.

%%%%%%%%%%%%%%%%%%%%%%%%%%%%%%%%%%%%%%%%%%%%%%%%%%%%%%%%%%%%%%%%%%%%%%
%   Sec. 5   Discussion and Summary
%%%%%%%%%%%%%%%%%%%%%%%%%%%%%%%%%%%%%%%%%%%%%%%%%%%%%%%%%%%%%%%%%%%%%%
\section{Discussion and Summary}

We have evaluated the quadrupole susceptibility for realistic values
of Coulomb interaction and spin-orbit coupling for Pr- and Gd-based
filled skutterudite compounds.
In particular, we have pointed out a possibility
that the softening of the elastic constant
can be observed for Gd compounds.

Concerning previous research on elastic properties of Gd compounds,
we notice the measurement of elastic constant in rare-earth hexaborides.
\cite{Nakamura}
No anomaly has been reported in GdB$_6$ due to the elastic constant
measurement, except for the signal at
a temperature of the antiferromagnetic phase transition.
In this sense, it may be difficult to observe actually
the softening of
elastic constant in Gd compounds with the cubic structure.
However, we believe that it is worthwhile to perform
more precise ultrasonic experiments in Gd compounds,
in particular, for Gd-based filled skutterudites.

Here we also remark that Eu$^{2+}$ also contains seven $f$ electrons.
Thus, we expect the appearance of
the softening of elastic constant in Eu compounds.\cite{Nakanishi}
Note, however, that valence fluctuation between divalent and trivalent
states is considered to be strong in europium.
It may be also difficult to detect the softening of elastic constant
in Eu compounds, but we hope that precise ultrasonic experiments
in Eu compounds will be done in near future.

In the evaluation of the quadrupole susceptibility,
we have not considered the competition among different types
of multipoles.
Since we have concentrated only on the measurement of
the elastic constant, the quadrupole susceptibility
has been evaluated.
However, for Gd compounds, we have frequently observed an
antiferromagnetic phase in the low-temperature region.
When the N\'eel temperature is so high, it may be difficult
to detect the softening of the elastic constant in the paramagnetic phase.
In such a case, it is interesting to consider the anisotropy
of magnetization in the antiferromagnetic phase
due to the effect of quadrupole degree of freedom.
It is one of future problems to estimate the extent of
the anisotropy of magnetization in Gd compounds.

If we intend to show explicitly the occurrence
of the antiferromagnetic phase,
it is necessary to perform the optimization of
multipole susceptibility.\cite{Hotta10,Hotta14}
Then, we conclude that the multipole state
with the largest eigenvalue of the susceptibility matrix
is realized when we decrease the temperature.
Such calculations have been actually performed for
the seven-orbital Anderson model with the use of
a numerical renormalization group technique.\cite{Hotta8}
For the case of Gd ion, it has been shown that
the dipole state exhibits the maximum eigenvalue
and the eigenstate of the second largest eigenvalue
is characterized by quadrupole.
As for the extension of the calculations to the periodic
systems,\cite{Hotta14} it is another future issue.

%%%% addition 2012/9/7
In this paper, we have considered the coupling between quadrupole
and the strain for the elastic constant.
However, from the symmetry viewpoint,
it is necessary to include also the effect of hexadecapoles.
In fact, for the explanation of the temperature dependence of
the elastic constant of PrMg$_3$ with $\Gamma_3$ doublet ground state,
a crucial role of hexadecapole has been emphasized.\cite{Araki}
Also in Gd-based compounds, such higher-order multipoles may play
some roles for the explanation of the behavior of elastic constant.
If we will successfully observe the softening of the elastic constant
in Gd-based filled skutterudites, it may be necessary to include
the effect of hexadecapole to explain precisely
the temperature dependence of the elastic constant.
It is also another future problem.
%%%%

Finally, we provide a comment on the normalization of
the multipole operator.
In the present paper, since we have considered
only the quadrupole susceptibility,
it is not necessary to pay attention to the normalization
of the multipole operator.
However, if we consider the diagonalization of the multipole
susceptibility matrix, it is inevitable to define
the normalization.
A natural way is to redefine each multipole operator
so as to satisfy
the orthonormal condition of
${\rm Tr} \{
{\hat T}_{\gamma}^{(k)} {\hat T}_{\gamma'}^{(k')} \}$
=$\delta_{kk'}\delta_{\gamma\gamma'}$.\cite{Kubo4}
Note again that we have not considered such redefinition
in the present paper.

In summary, we have found that the $j$-$j$ coupling component
significantly appears in the ground-state wavefunction
for $n$=7 with $J$=$7/2$ even for realistic values of
Coulomb interaction and spin-orbit coupling.
With the use of the CEF parameters determined from PrOs$_4$Sb$_{12}$,
we have evaluated the quadrupole susceptibility for
Gd-based filled skutterudites.
We have emphasized that the softening of the elastic
constant can be detected in Gd compounds.
We expect that ultrasonic experiments will be performed
in Gd-based filled skutterudite compounds in a future.

\section*{Acknowledgement}

We thank Y. Aoki, Y. Nakanishi, Y. Nemoto, H. Sato, T. Goto,
T. Yanagisawa, and Y. Yoshizawa for discussion and comments.
This work has been supported by a Grant-in-Aid for Scientific Research
on Innovative Areas ``Heavy Electrons''
(No. 20102008) of The Ministry of Education, Culture, Sports,
Science, and Technology, Japan. 
The computation in this work has been done using the facilities
of the Supercomputer Center of Institute for Solid State Physics,
University of Tokyo.

%%%%%%%%%%%%%%%%%%%%%%%%%%%%%%%%%%%%%%%%%%%%%%%%%%%%%%%%%%%%%%%%%%%%%%
%   References
%%%%%%%%%%%%%%%%%%%%%%%%%%%%%%%%%%%%%%%%%%%%%%%%%%%%%%%%%%%%%%%%%%%%%%


\begin{thebibliography}{99}

%-------------------- Sec. 1 -----------------------------

\bibitem{Hotta1}
T. Hotta:
Rep. Prog. Phys. {\bf 69} (2006) 2061.

\bibitem{Review2}
Y. Kuramoto, H. Kusunose, and A. Kiss:
J. Phys. Soc. Jpn. {\bf 78} (2009) 072001.

\bibitem{Review3}
P. Santini, S. Carretta, G. Amoretti, R. Caciuffo, N. Magnani,
and G. H. Lander:
Rev. Mod. Phys. {\bf 81} (2009) 807.

\bibitem{skut}
H. Sato, H. Sugawara, Y. Aoki, and H. Harima:
{\it Handbook of Magnetic Materials} Volume 18,
ed. K. H. J. Buschow, pp. 1-110, Elsevier, Amsterdam, 2009.

\bibitem{Kohgi}
M. Kohgi, K. Iwasa, M. Nakajima, N. Metoki, S. Araki, N. Bernhoeft,
J.-M. Mignot, A. Gukasov, H. Sato, Y. Aoki and H. Sugawara:
J. Phys. Soc. Jpn. {\bf 72} (2003) 1002.

\bibitem{Kuwahara}
K. Kuwahara, K. Iwasa, M. Kohgi, K. Kaneko, S. Araki, N. Metoki,
H. Sugawara, Y. Aoki and H. Sato:
J. Phys. Soc. Jpn. {\bf 73} (2004) 1438.

\bibitem{Goremychkin}
E. A. Goremychkin, R. Osborn, E. D. Bauer, M. B. Maple,
N. A. Frederick, W. M. Yuhasz, F. M. Woodward and J. W. Lynn:
Phys. Rev. Lett. {\bf 93} (2004) 157003.

%------------------- Sec. 2 --------------------------

\bibitem{Slater}
J. C. Slater:
{\it Quantum Theory of Atomic Structure},
(McGraw-Hill, New York, 1960).

\bibitem{Hutchings}
M. T. Hutchings:
Solid State Phys. {\bf 16} (1964) 227.

\bibitem{Takegahara}
K. Takegahara, H. Harima and A. Yanase:
J. Phys. Soc. Jpn. {\bf 70} (2001) 1190.
%{\it ibid.} {\bf 70} (2001) 3468;
%{\it ibid.} {\bf 71} (2002) 372.

\bibitem{LLW}
K. R. Lea, M. J. M. Leask and W. P. Wolf:
J. Phys. Chem. Solids {\bf 23} (1962) 1381.

%%% NpO2 micro theory
\bibitem{Kubo1}
K. Kubo and T. Hotta:
Phys. Rev. B {\bf 71} (2005) 140404(R).

\bibitem{Kubo2}
K. Kubo and T. Hotta:
Phys. Rev. B {\bf 72} (2005) 132411.

\bibitem{Kubo3}
K. Kubo and T. Hotta:
Phys. Rev. B {\bf 72} (2005) 144401.

%%% f2 mag. fluctuation
\bibitem{Hotta2}
T. Hotta:
Phys. Rev. Lett. {\bf 94} (2005) 067003.

%%% LS vs. j-j
\bibitem{Hotta3}
T. Hotta:
J. Phys. Soc. Jpn. {\bf 74} (2005) 1275.

%%% multipole sus.
\bibitem{Hotta4}
T. Hotta:
J. Phys. Soc. Jpn. {\bf 74} (2005) 2425.

%%% modified j-j
\bibitem{Hotta5}
T. Hotta and H. Harima:
J. Phys. Soc. Jpn. {\bf 75} (2006) 124711.

%%% Nd
\bibitem{Hotta6}
T. Hotta:
J. Magn. Magn. Mater. {\bf 310} (2007) 1691.

%%% Sm
\bibitem{Hotta7}
T. Hotta:
J. Phys. Soc. Jpn. {\bf 76} (2007) 034713.

%%% heavy rare-earth
\bibitem{Hotta8}
T. Hotta:
J. Phys. Soc. Jpn. {\bf 76} (2007) 083705.

\bibitem{Hotta9}
T. Hotta:
J. Phys. Soc. Jpn. {\bf 77} (2008) 074716.

\bibitem{Hotta10}
T. Hotta:
{\it Proc. Int. Conf. New Quantum Phenomena in Skutterudite and
Related Systems (Skutterudite 2007)},
J. Phys. Soc. Jpn. {\bf 77} (2008) Suppl. A, p. 96.

\bibitem{Hotta11}
T. Hotta:
J. Phys.: Conf. Ser. {\bf 150} (2009) 042061.

\bibitem{Hotta12}
T. Hotta:
Phys. Rev. B {\bf 80} (2009) 024408.

\bibitem{Hotta13}
T. Hotta:
J. Phys. Soc. Jpn. {\bf 79} (2010) 094705.

\bibitem{Hotta14}
T. Hotta:
Phys. Res. Int. {\bf 2012} (2012) 762798.

\bibitem{Inui}
T. Inui, Y. Tanabe and Y. Onodera:
{\it Group Theory and Its Applications in Physics},
(Springer, Berlin, 1996).

\bibitem{StevensOP}
K. W. H. Stevens:
Proc. Phys. Soc. A{\bf 65} (1952) 209.

\bibitem{Shiina1}
R. Shiina, H. Shiba and P. Thalmeier:
J. Phys. Soc. Jpn. {\bf 66} (1997) 1741.

\bibitem{Shiina2}
R. Shiina: J. Phys. Soc. Jpn. {\bf 73} (2004) 2257.

%%% Spin-orbit coupling
\bibitem{spin-orbit}
S. H\"ufner:
{\it Optical Spectra of Transparent Rare Earth Compounds},
(Academic Press, New York, 1978).

\bibitem{Goto1}
T. Goto, Y. Nemoto, K. Sakai, K. Onuki, T. Yamaguchi, M. Akatsu,
T. Yanagisawa, H. Sugawara, and H. Sato:
Physica B {\bf 359-361} (2005) 822.

\bibitem{Goto2}
T. Goto, Y. Nemoto, K. Onuki, K. Sakai, T. Yamaguchi, M. Akatsu,
T. Yanagisawa, H. Sugawara, and H. Sato:
J. Phys. Soc. Jpn. {\bf 74} (2005) 263.

%%% GdB6
\bibitem{Nakamura}
S. Nakamura, T. Goto, S. Kunii, K. Iwashita, and A. Tamaki:
J. Phys. Soc. Jpn. {\bf 63} (1994) 623.

\bibitem{Nakanishi}
Y. Nakanishi:
private communications.

%%% PrMg3
\bibitem{Araki}
K. Araki, Y. Nemoto, M. Akatsu, S. Jumonji, T. Goto, H. S. Suzuki,
H. Tanida, and S. Takagi: Phys. Rev. B {\bf 84} (2011) 045110.

%%% orthonormal condition
\bibitem{Kubo4}
K. Kubo and T. Hotta:
J. Phys. Soc. Jpn. {\bf 75} (2006) 013702.

\end{thebibliography}
\end{document}